# An 8-Gs/s 12-Bit TIADC System with Real-Time Broadband Mismatch Error Correction

Lei Zhao, *Member, IEEE,* Zouyi Jiang, Ruoshi Dong, Zhe Cao, *Member, IEEE*, Xingshun Gao, Boyu Cheng, Jiadong Hu, Shubin Liu, *Member, IEEE,* Qi An, *Member, IEEE*

*Abstract*—High sampling speed can be achieved using multiple Analog-to-Digital Converters (ADCs) based on the Time-Interleaving A/D Conversion (TIADC) technique. Various types of methods were proposed to correct the mismatch errors among parallel ADC channels in TIADC systems, which would deteriorate the system performance. Traditional correction methods based on digital signal processing have good performance, however often only for input signals limited in a narrow frequency band. In this paper, we present our recent work on design of an 8-Gsps 12-bit TIADC system and implementation of real-time mismatch correction algorithms in FPGA devices, over a broad band of input signal frequencies. Tests were also conducted to evaluate the systems performance, and the results indicate that the Effective Number of Bits (ENOB) is enhanced to be better than 8.5 bits (<800 MHz) and 8 bits from 800 MHz to 1.6 GHz after correction, almost the same with that of the ADC chip employed.

## I. Introduction

WAVEFORM digitization is a preferable solution in physics to obtain the most detailed information from the signals out of detectors, and has thus been employed in many physics experiments [1]-[11]. With the development of electronics, especially ASIC design on Analog-to-Digital Converters (ADCs), sampling speed has been increasing. With the Time Interleaved A/D Conversion (TIADC) technique, the system sampling speed can be greatly enhanced beyond single ADC ASIC's capability [8]-[10], which makes the study in this direction a research hot spot.

Manuscript received Jul. 6, 2018. This work was supported in part by the National Natural Science Foundation of China under Grant 11675173, in part by the Knowledge Innovation Program of the Chinese Academy of Sciences under Grant KJCX2-YW-N27, and in part by the CAS Center for Excellence in Particle Physics (CCEPP).
The authors are with the State Key Laboratory of Particle Detection and Electronics, University of Science and Technology of China, Hefei 230026, China and Department of Modern Physics, University of Science and Technology of China, Hefei 230026, China (Corresponding author: Qi An, e-mail: anqi@ustc.edu.cn).
© 2018 IEEE. Accepted version for publication by IEEE. Digital Object Identifier 10.1109/TNS.2018.2878875

In TIADC systems, the basic idea to use multiple ADC channels working in parallel, while the phases of the clock signals for multiple channels are shifted by a predetermined interval. There exist inevitable mismatch errors among different ADC channels in TIADC systems [12]-[14]. Therefore, methods of mismatch error correction is an important research domain.

Efforts have been devoted to address this issue, and many approaches were proposed. In general, these correction methods can be categorized into analog adjustment and digital correction. As for the former, a monitor channel is usually designed to detect the mismatch errors, and then the multiple ADC channels can be adjusted to reduce the mismatch among them [15][16]. However, since additional analog circuits are used (susceptible to noise, interference, etc.), the correction resolution is limited; for example, in [16] a sampling speed of 12.8 Gsps is achieved, but the Effective Number of Bits (ENOB) is only around 4.6 bits (32 channels of 7-bit 400-Msps ADCs are interleaved) with the analog adjustment method applied. As for digital correction, it can be further categorized into background and foreground correction. The former is also addressed as self-adaptive method [17]-[21], and it does not require a calibration process before use, but the circuit is quite complex and the input signal in real application must be Wide-Sense Stationary (WSS) [22], which limits the application of this correction method. The foreground correction method includes the interpolation method, the method based on fractional delay filters [23], and the perfect reconstruction method [24]-[28]. Compared with the first and second methods, which require complex logic design and over-sampling, respectively, the perfect reconstruction correction method is a favorable choice. However, in traditional correction methods, good performance can be achieved only within a narrow input signal bandwidth, while the signals in physics experiments and many other domains are mostly wide band signals. Therefore, wide band signal mismatch correction becomes the key issue in this domain.

This paper presents the design of an 8-Gsps 12-bit digitizer based on the TIADC method. To address the mismatch among ADC channels, a real-time wide band correction algorithm integrated in a Field Programmable Gate Array (FPGA) device has been designed.

With the real-time mismatch correction algorithm, this digitizer achieves an ENOB of better than 8.5 bits (<800 MHz) and 8 bits from 800 MHz to 1.6 GHz. The technical details are discussed in the following sections.

## II. SYSTEM ARCHITECTURE

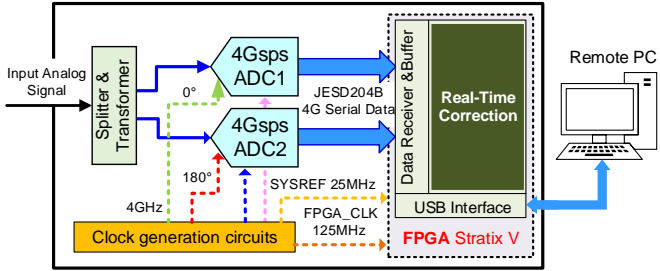

Fig. 1. Hardware structure of the 8 Gsps 12 bit TIADC system.

As shown in Fig. 1, the input signal is split into two paths, and sampled simultaneously by two 12 bit 4 Gsps ADCs (ADC12J4000 from Texas Instruments Corporation [29]) working in parallel. By adjusting the phase difference of the sampling clocks for these two ADCs to 180°, an equivalent overall sampling speed up to 8 Gsps can be achieved. The data interface between the ADCs and FPGA is based on the JESD204B standard [30], and thus the clock generation circuits also need to output 25 MHz synchronous clocks for the ADC readout. To correct the gain, offset and time skew errors among these two ADC channels, a real-time correction algorithm is designed and implemented in an FPGA device 5SGSMD6K2F40I2 from ALTERA Corporation.

## III. ANALOG-TO-DIGITAL CONVERSION CIRCUITS

*A. Structure of A/D Conversion circuits*

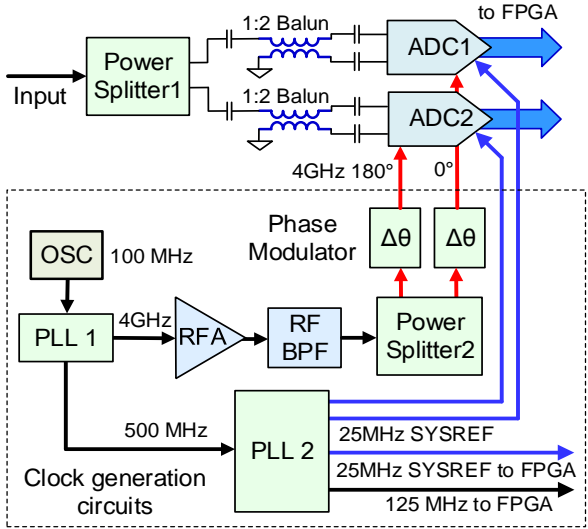

Fig. 2. Structure of the A/D conversion circuits.

Fig. 2 shows the structure of the A/D conversion circuits. As mentioned above, two ADCs are used to sample the input data in parallel, in order to achieve an overall sampling speed of 8 Gsps. The analog input signal is first fed to a power splitter (RPS-2-30+ [31] from Mini-Circuits corporation) and then its two output signals are converted to differential pairs through balun transformers (TCM2-43X+ [32] from Mini-Circuits Corporation) before A/D conversion.

*B. Clock Generator Circuits*

We use a high quality oscillator CCPD575 [33] from CRYSTEK corporation as the clock source of 100 MHz, and employ the PLL LMX2582 [34] (from Texas Instruments Corporation, marked as "PLL1" in Fig. 2) to generate a 4 GHz clock used for A/D Conversion and another 500 MHz clock for data transfer between ADC and FPGA. To further enhance the quality of the 4 GHz clock, we use a cascade of Radio Frequency Amplifier (RFA) MAAL-011078 [35] and a Band Pass Filter (BPF) BFCN-4100+ [36] (with pass band is from 3.7 GHz to 4.5 GHz) to suppress the phase noise out of band and achieve a proper amplitude of the clock signal. Then the clock is further split into two paths and their phases are tuned by two digitally controlled phase modulators to finally obtain two sampling clocks with a phase shift of 180° between them. To estimate the quality of the sampling clocks, simulations were conducted. Since the clock signal from the PLL is further enhanced by the RF circuits in Fig. 2, firstly we estimate the amplitude frequency response of the RF circuits. Shown in Fig. 3 is the S21 parameter (i.e. the amplitude frequency response) of the RF circuits (marked in red color), and the blue curve refers to the S21 of the overall circuits from PLL output to the ADC clock input. Second, we conducted simulations to obtain the phase noise of the PLL (marked in blue color in Fig. 4) using PLLatinumSim tool from Texas Instruments Corporation. Finally, combining the above information, we can get the phase noise curve of the clock signal fed to the ADC, as marked in red color in Fig. 4, and the estimated jitter of the ADC clock signal is around 82.62 fs.

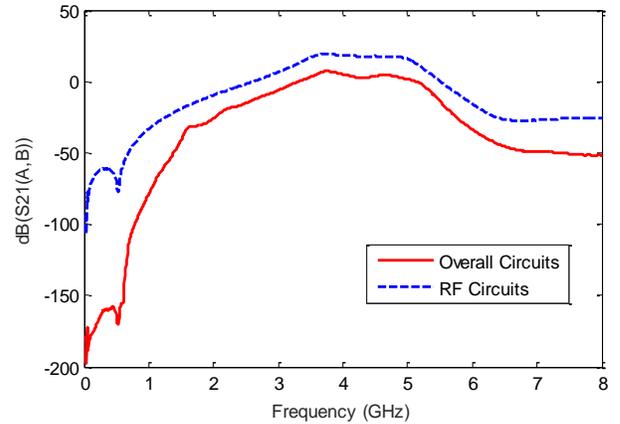

Fig. 3. Amplitude frequency response of the RF circuits.

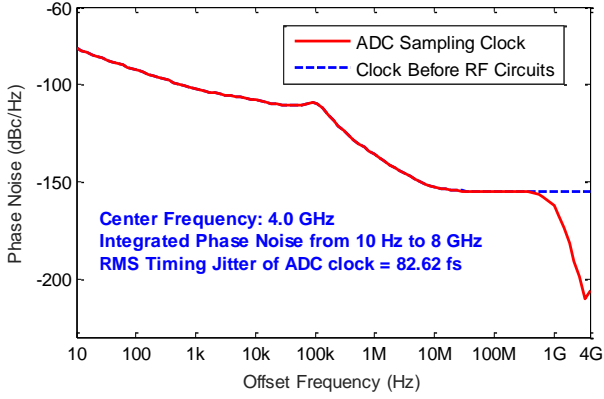

Fig. 4. Phase noise simulation results.

According to the relationship between Signal-to-Noise Ratio (SNR) and the sampling clock jitter, as in

$$SNR = -20\log 2\pi f_{in} t_{jitter} \quad (1),$$

the SNR with different input signal frequencies caused by the sampling clock can be calculated, and the results are listed in TABLE I.

TABLE I. INFLUENCE OF THE CLOCK JITTER

| Input Frequency (MHz) | SNR (dB) | ENOB (bits) |
|---|---|---|
| 100 | 85.7 | 13.9 |
| 500 | 71.7 | 11.6 |
| 900 | 66.6 | 10.8 |
| 1300 | 63.4 | 10.2 |
| 1600 | 61.6 | 9.9 |
| 2800 | 56.8 | 9.1 |

The above SNR and the corresponding ENOB results are good enough for this TIADC design (for example, the ENOB due to clock is 9.1 bits, much better than that of a single ADC, an ENOB of 7.7 bits @ 2.8 GHz input frequency [29]).

### C. Data Transfer Interface

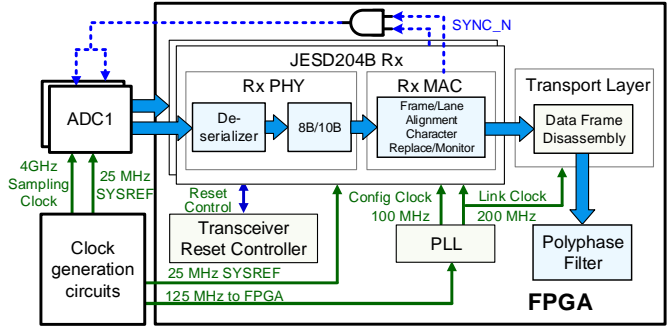

Fig. 5. Data transfer between the ADCs and FPGA.

The data speed from the ADCs to the FPGA is up to 96 Gbps, which is quite high. To achieve a high data rate up to 48 Gbps per ADC chip, a JESD204B interface is employed in the ADCs. Shown in Fig. 5 is the data interface between the ADCs and the FPGA. According to the JESD204B application requirement, 25 MHz synchronous clocks (marked as "SYSREF" in Fig. 2 and Fig. 5) need to be provided for the ADC and FPGA simultaneously. In addition, a 125 MHz system clock is sent to the FPGA to generate the required 100 MHz and 200 MHz link clocks (marked as "Config Clock" and "Link Clock" in Fig. 5). To receive the data from the ADC, the JESD204B IP Core [37] is employed within the FPGA, which consists of the PHY and MAC layers. Logic is also designed to disassemble the ADC data frame and repackage the data to accommodate the following mismatch correction processing logic. The flag signals from the two JESD204B IP cores are also used to generate a "SYNC_N" signal to synchronize the data streams of the two ADCs.

## IV. REAL-TIME CORRECTION BASED OVER WIDE FREQUENCY BAND

### A. Correction Algorithm

As mentioned above, good correction results can be achieved for narrow frequency band input signals using the previously mentioned correction methods. However, in the current application, the input pulses usually contain energy over a wide frequency range. In this paper, we implemented real time correction methods over a wide band in the FPGA device.

When considering wide band input signal, we should notice that among the three mismatch errors (i.e. offset, gain, and time skew), gain and time skew mismatches both vary with frequency. As for time skew mismatch, it is caused not only by the delay difference among clock signals for multiple ADCs but also by the input signal delay difference, which actually corresponds to the phase response of the front analog circuits and of course changes with frequency.

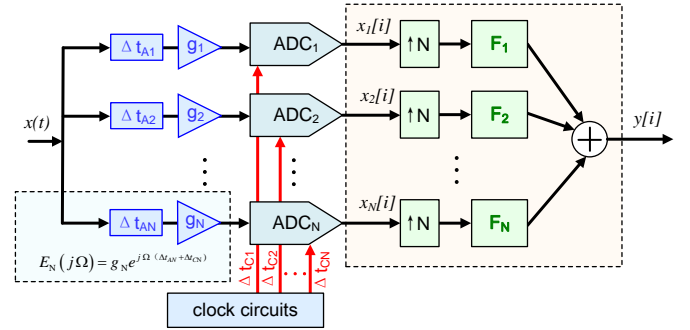

Fig. 6. TIADC structure and mismatch errors.

Fig. 6 illustrates the working principle of the TIADC system and indicates mismatch errors due to different factors. The offset mismatch is a constant value which can be easily corrected by subtracting this offset from the digitized signal waveform. The gain mismatch value can be expressed as a function of the input signal frequency $\Omega$, i.e. the mismatch value changes with $\Omega$. Only in a comparatively low frequency range, the gain mismatch could be considered a constant value. The time skew error comes from two sources. The first is the mismatch among the clock phases received by different ADCs, which can be considered as a constant value. The other is the mismatch of the input analog signal delay before A/D conversion, which also influences the sampling time point on the signal waveform and changes with $\Omega$. Especially in very high speed TIADC systems, small time skew errors would influence the system performance significantly. This means much attention should be paid on the issue of parameter dependence of $\Omega$.

In traditional methods, good correction effect can be achieved, but the coefficients in the correction algorithms are fixed, which means that the correction performs well only within a narrow frequency band. For example, in our previous work [38], correction was performed on a 14-bit 1.6-Gsps TIADC, with system performance enhanced over an input frequency range from 5 MHz to 600 MHz. However, for each input frequency point in the test, we had to change the coefficients of the FIR filter used for correction. In actual application in physics experiments, the input signals are usually pulses, the energy of which are distributed over a wide frequency band. In this case, we need to correct the mismatch errors with only one single set of filter coefficients.

As shown in Fig. 6, the mismatch gain and time skew mismatch errors can be expressed as (taking Channel N for example):

$$E_N(j\Omega) = g_N e^{j\Omega \Delta t_{TN}} = g_N e^{j\Omega(\Delta t_{AN} + \Delta t_{CN})} \quad (2).$$

Then the final output data after A/D conversion can be calculated as in [39]:

$$Y(e^{j\omega}) = \frac{1}{T_s} \sum_{i=-\infty}^{\infty} \left[ \frac{1}{N} \sum_{n=1}^{N} E_n \left( \frac{j(\omega - 2\pi i/N)}{T_s} \right) \right] \cdot X \left( \frac{j(\omega - 2\pi i/N)}{T_s} \right) \quad (3),$$

where $\omega = \Omega T_S$, $T_S$ is the period of the sampling clock, $g_n$ is the gain of Channel n, $\Delta t_{Cn}$ is the clock skew mismatch of Channel n compared with the first ADC channel, $\Delta t_{An}$ is the mismatch error of the input signal delay caused by the analog front end before the ADC. From (3) we can observe the influence caused by the gain and time skew error.

With a correction algorithm applied, the output data can now be expressed as in

$$Y(e^{j\omega}) = \frac{1}{T_s} \sum_{i=-\infty}^{+\infty} X \left( j\frac{\omega}{T_s} - j\frac{2\pi i}{NT_s} \right) H_i \left( j\frac{\omega}{T_s} - j\frac{2\pi i}{NT_s} \right) \quad (4).$$

where $H_i$ contains the effects of both the mismatch errors and the correction filters.

After A/D conversion, only the signal energy within the first Nyquist zone can be observed, i.e. in the range of $[-\pi, \pi]$ normalized by $T_S$. In this range, the output data are

$$Y^S(e^{j\omega}) = \frac{1}{T_s} \sum_{i=0}^{N-1} X^s \left( j\frac{\omega}{T_s} - j\frac{2\pi i}{NT_s} \right) H_i^s \left( j\frac{\omega}{T_s} - j\frac{2\pi i}{NT_s} \right) \quad (5).$$

And $H_i$ is expressed as

$$H_i^s(e^{j\omega}) = \frac{1}{N} \sum_{n=1}^{N} F_n(e^{j\omega}) \cdot E_n \left( \frac{j(\omega - 2\pi i/N)}{T_s} \right) \quad (6),$$

where $F_n$ corresponds to the filter in Channel n used for correction.

For a perfect correction, it is required that

$$H_i^s(e^{j\omega}) = \begin{cases} g_{fix} e^{-j\omega \Delta t_{fix}}, & i = 0 \\ 0, & i = 1..., N-1 \end{cases} \quad (7),$$

which means that after correction the output $Y$ is the perfect reconstruction of the input $X$ with a fixed gain of $g_{fix}$ and a fixed delay $\Delta t_{fix}$, both independent of frequency, and no interference from aliasing exists.

In traditional methods, the basic idea is to solve the expression of $F_n$ according to (6) and (7), and calculate the filter coefficients as

$$f_n[i] = \frac{1}{2\pi} \int_{-\pi}^{\pi} F_n(e^{j\omega}) e^{j\omega i} d\omega, i=1,...R \quad (8),$$

where $R$ is the order of the filter.

The above method works well when the mismatch parameters $g_n$, $\Delta t_{An}$, and $\Delta t_{Cn}$ are fixed values. This happens within a narrow input frequency band, but in a wide frequency band the mismatch parameters would change, which means (6) and (7) do not exist and thus $F_n$ can not be solved, i.e. the above method does not work anymore.

To address this issue, our idea is to directly calculate numerically the filter coefficients instead of solving the expression of $F_n$.

Since the time skew error $\Delta t_{Tn} = \Delta t_{An} + \Delta t_{Cn}$ and gain $g_n$ varies with input signal frequency, in the first step we calibrate these mismatch errors on multiple frequency points ($\omega_1, \ldots \omega_K$) within a wide band. And then from (6) and (7) we can calculate $F_n(\omega_k)$ at such frequency points as

$$\begin{bmatrix} E_{1,k}\left(\frac{j\omega_k}{T_s}\right) & \cdots & E_{N,k}\left(\frac{j\omega_k}{T_s}\right) \\ E_{1,k}\left(\frac{j(\omega_k - 2\pi/N)}{T_s}\right) & \cdots & E_{N,k}\left(\frac{j(\omega_k - 2\pi/N)}{T_s}\right) \\ \vdots & \ddots & \vdots \\ E_{1,k}\left(\frac{j(\omega_k - 2\pi(N-1)/N)}{T_s}\right) & \cdots & E_{N,k}\left(\frac{j(\omega_k - 2\pi(N-1)/N)}{T_s}\right) \end{bmatrix} \cdot \begin{bmatrix} F_1(e^{j\omega_k}) \\ F_2(e^{j\omega_k}) \\ \vdots \\ F_N(e^{j\omega_k}) \end{bmatrix} = \begin{bmatrix} g_{fix} e^{-j\omega \Delta t_{fix}} \\ 0 \\ \vdots \\ 0 \end{bmatrix} \quad (9),$$

where $E_{n,k}$ can be calculated using the calibration results of $\Delta t_{Tn}(\omega_k)$ and $g_n(\omega_k)$. From (9), $F_0, \ldots F_{N-1}$ at frequency point $\omega_k$ can be obtained. By sweeping the frequency from $\omega_1$ to $\omega_K$, we can finally obtain all $F_n(\omega_k)$ that we need. Actually in real applications, we can obtain the mismatch parameters on one set of frequency points, and calculate the parameters on other frequency points thought fitting and interpolation if we need.

According to (9), we can further calculate the filter coefficients $f_n[i]$ with approximation as

$$f_n[i] = \frac{1}{2\pi} \int_{-\pi}^{\pi} F_n(e^{j\omega}) e^{j\omega i} d\omega \approx \frac{1}{K} \sum_{k=1}^{K} F_n(e^{j\omega_k}) e^{j\omega_k i} \quad (10).$$

### B. Simulations

To evaluate the performance of the above correction method, we conducted a series of simulations.

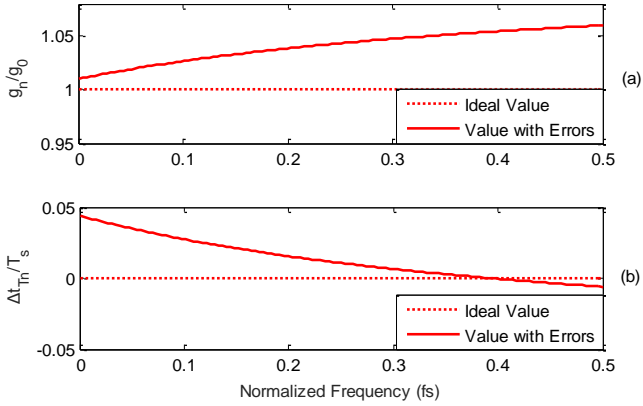

Fig. 7. Mismatch errors used in the simulations. (a) gain mismatch; (b) time skew mismatch.

Shown in Fig. 7 is the mismatch errors used in the simulation, in which both the gain ($g_n$) and time skew ($\Delta t_{Tn}$) errors change with input signal frequency. In order to implement the algorithm in an FPGA, two key parameters are of great concern: the filter order and bit width of the filter coefficients, which directly determines the resource consumption in the FPGA.

First, we change the filter order and sweep from 40 to 160 and observe the correction performance variation, as shown in Fig. 8.

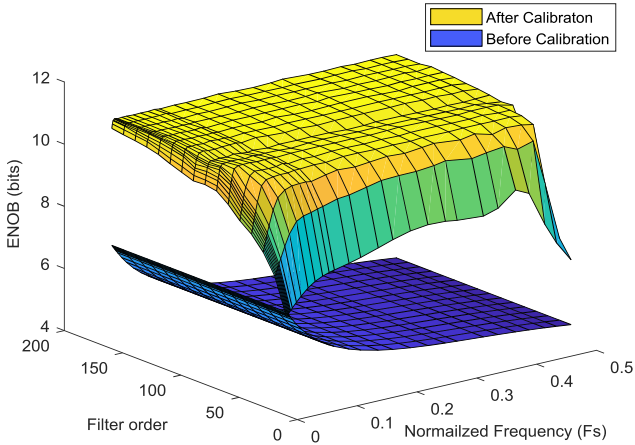

Fig. 8. Correction effects vs filter order.

Fig. 8 indicates that with a larger filter order, a better effect can be achieved. With a filter order of ≥ 80, the ENOB is effectively enhanced over a wide frequency range in the first Nyquist zone.

The next simulation is for evaluation of the bit width of the filter coefficients. We sweep this parameter from 12 bits to 30 bits with a fixed filter order of 80.

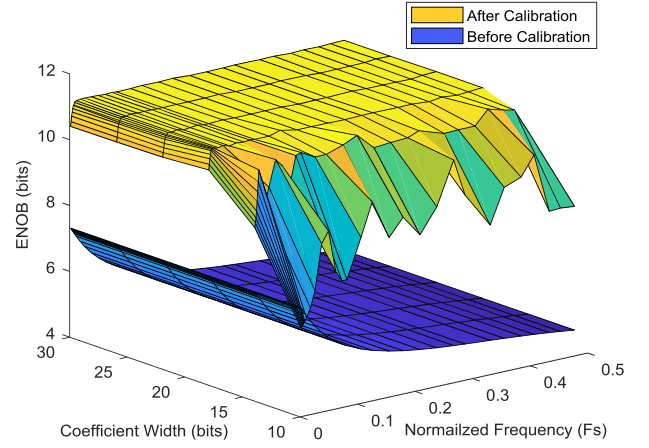

Fig. 9. Correction effects vs filter coefficient bit width.

As shown in Fig. 9, when the coefficient width is bigger than 16 bits, the correction effect is good enough.

As mentioned above, the most comprehensive information of the detector output signal can be obtained through waveform digitization in physics experiments. Especially in some experiments, the digitized waveform needs to be used for Pulse Shape Discrimination (PSD) [11][40]-[42]. Better performance of the TIADC would allow more precise observation of the detector output signal. Even if in the applications only the time and charge are needed, the enhanced quality of the TIADC system after mismatch error correction results in a better resolution, especially for time measurement when the waveform varies.

We also conducted simulations to estimate the correction effect. In the simulations, we use a typical pulse, with a rise and trailing edge of 1 ns and 2.8 ns, respectively, as shown in Fig. 10. In Fig. 10 (b), it can be observed that the waveform distortion due to mismatch errors can be effectively decreased by the correction process.

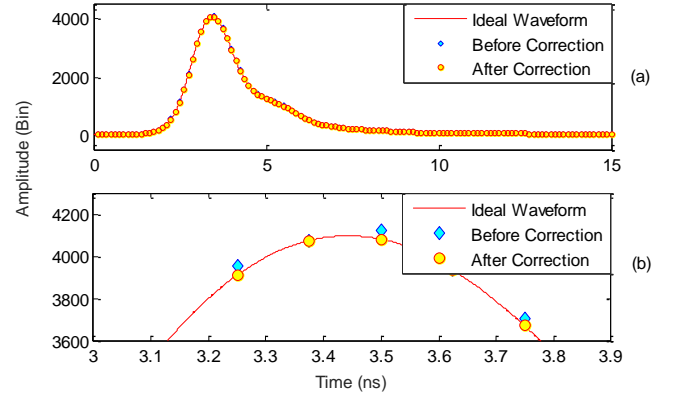

Fig. 10. Waveform used in time resolution simulation. (a) signal waveform; (b) partial waveform nearby signal peak zoomed in.

In the next step, we change the rise and trailing edge of the waveform, stretching them by a ratio that is a random value distributed in a specific range. At the same time we also change the peak value of the waveform to keep the waveform area constant. Then we can simulate the time resolution when the signal waveform changes its shape. Next, we conduct fitting on the digitized waveform and calculate the time information of each pulse. With a certain number of waveforms obtained, we can finally get the time resolution through statistical analysis.

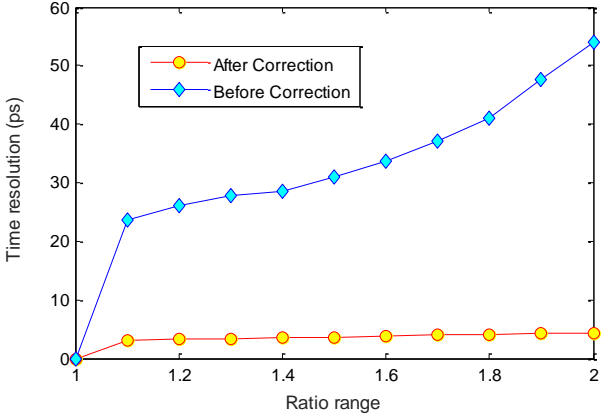

Fig. 11. Time resolution simulation results before and after correction.

Fig. 11 shows the time resolution before and after correction with different ratio ranges (for example, when the ratio range is 1.2, it means that the rise edge is stretched by a ratio varying from 1 to 1.2, the same with the trailing edge). Fig. 11 indicates that the time resolution can be effectively enhanced by the correction process.

### C. Real-Time Correction Algorithm Implementation

After the above work, we further implement the correction algorithm in the FPGA device. Considering the high speed of the data stream, much attention is paid on transforming the correction algorithm to a parallel structure.

First, we separate the data streams of the two ADCs into 40 groups. The detailed process is as follows. Shown in Fig. 12 (a) is the data frame from the ADC. Each ADC outputs 48 Gbps (4 Gsps with a 12 bit data width) data streams which are distributed into 8 lanes. As for each lane, there are 8 bytes in each frame with 4 tail bits fixed to '0'. As mentioned above, we use the JESD204B IP core in the FPGA to receive the ADC data, and the data streams out of this core are of 480-bit width (as shown in Fig. 12 (b)). This 480 bit wide frame is output in two phases marked as "Link Clock Period 1" and "Link Clock Period 2" in Fig. 12. In each phase, half of the frame (240-bit width, e.g. the left part of Fig. 12 (b) separated by the dotted line) is fed out. To extract the 12 bit wide data and align them into 40 channels, we repackage the data frame using combinational logic in a pipelined structure, and the data from ADC1 are assigned to Ch0, 2, … 38 with ADC2 output data assigned to Ch1, 3, … 39, as shown in Fig. 12.

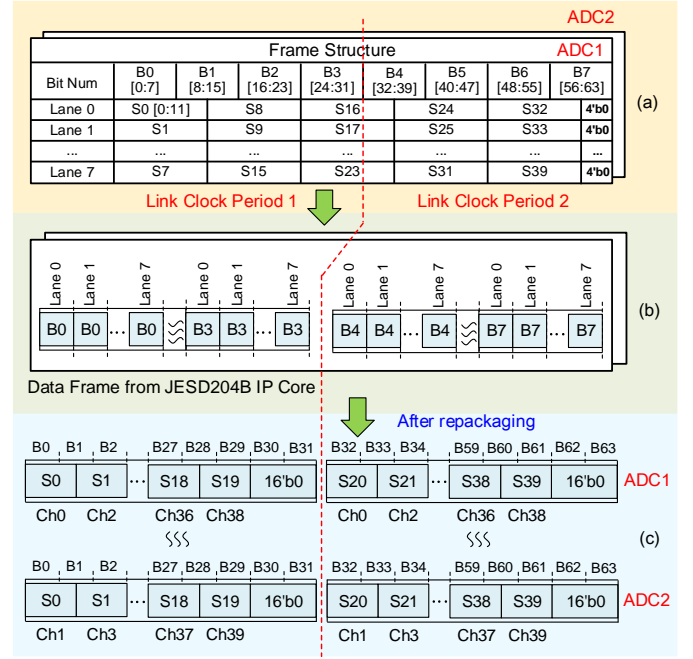

Fig. 12. Data extraction and repackaging. (a) ADC output data frame; (b) data frame from JESD204B IP Core; (c) data streams after repackaging.

With these 40 data streams as input for correction (marked as Ch0, Ch1…, and Ch39 in Fig. 12 and Fig. 13), the processing speed is thus decreased from 8 Gsps to 200 Msps. The structure of the correction algorithm is shown in Fig. 13. Each data stream is first corrected with the offset value, then up-sampled by a factor 40, and passed through an 80-order FIR filter. The processed data streams are then delayed for several clock periods and finally summed together to reconstruct the corrected digital waveform.

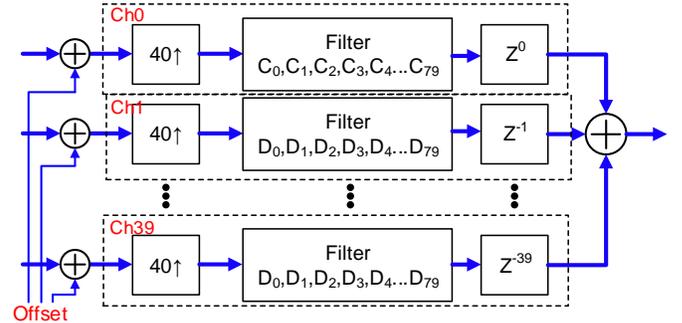

Fig. 13. Structure of the correction algorithm.

As for the design of the correction filter for this TIADC, since 40 parallel channels are needed to achieve real time processing of the 96 Gbps data stream, if we implement this parallel algorithm using the previous method based on interpolation filter [38], the signal process would be expressed as in

$$\begin{bmatrix} Y_0(z) \\ Y_1(z) \\ \vdots \\ Y_{39}(z) \end{bmatrix} = \begin{bmatrix} F_{0,0}(z) & F_{0,1}(z) & \cdots & F_{0,39}(z) \\ F_{1,0}(z) & F_{1,1}(z) & \cdots & F_{1,39}(z) \\ \vdots & \vdots & \ddots & \vdots \\ F_{39,0}(z) & F_{39,1}(z) & \cdots & F_{39,39}(z) \end{bmatrix} \bullet \begin{bmatrix} X_0(z) \\ X_1(z) \\ \vdots \\ X_{39}(z) \end{bmatrix} \quad (11).$$

According to (11), it means that for the data stream in each channel $X_i$, the data need to be further fanned out to 40 streams.

It will cause complex signal routing within the FPGA, which would make the time constraints difficult to be met in high speed situations. To address this issue, we propose to reorganize the process into two steps. First, we fan out each $X_i$ to five streams, and then correct the data in parallel, as in

$$\begin{bmatrix} Y_0(z) \\ Y_5(z) \\ \vdots \\ Y_{35}(z) \\ Y_1(z) \\ Y_6(z) \\ \vdots \\ Y_{36}(z) \\ \vdots \\ Y_4(z) \\ Y_9(z) \\ \vdots \\ Y_{39}(z) \end{bmatrix} = \begin{bmatrix} F^0_{8\times40} \\ 0_{32\times40} \end{bmatrix} \cdot \begin{bmatrix} X_0(z) \\ X_1(z) \\ \vdots \\ X_{39}(z) \end{bmatrix} + \begin{bmatrix} 0_{8\times40} \\ F^1_{8\times40} \\ 0_{24\times40} \end{bmatrix} \cdot \begin{bmatrix} X_0(z) \\ X_1(z) \\ \vdots \\ X_{39}(z) \end{bmatrix} + \cdots + \begin{bmatrix} 0_{32\times40} \\ F^4_{8\times40} \end{bmatrix} \cdot \begin{bmatrix} X_0(z) \\ X_1(z) \\ \vdots \\ X_{39}(z) \end{bmatrix} \quad (12),$$

$$F^k_{8\times40} = \begin{bmatrix} F_{k,0}(z) & F_{k,1}(z) & \cdots & F_{k,39}(z) \\ F_{k+5,0}(z) & F_{k+5,1}(z) & \cdots & F_{k+5,39}(z) \\ \vdots & \vdots & \ddots & \vdots \\ F_{k+35,0}(z) & F_{k+35,1}(z) & \cdots & F_{k+35,39}(z) \end{bmatrix} \quad (13),$$

where k=0, 1, 2, 3, 4, and each value of k corresponds to one data stream after fan-out. With this method, the matrix in (11) of 40 × 40 is transformed to five matrices of 8 × 40. The corresponding structure of the processing logic is shown in Fig.14.

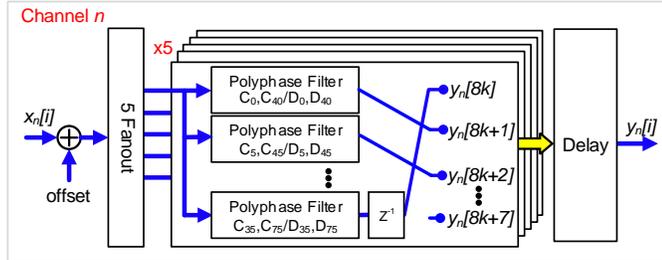

Fig. 14. Parallel structure of each channel.

Through the above steps, the complex correction algorithm implemented in an FPGA is feasible, and we actually designed this algorithm in one FPGA device 5SGSMD6K2F40I2, based on which real-time correction for an 8-Gsps 12-bit TIADC is achieved. Shown in TABLE II is the resource consumption of the overall logic, including the correction algorithm and data interface.

TABLE II  RESOURCE CONSUMPTION

| Resource | LUT | DSP | XCVR | FF | Block Memory |
|---|---|---|---|---|---|
| Used | 50362 | 1280 | 16 | 111871 | 16949248 |
| Utilization Ratio | 23% | 72% | 44% | 12% | 36% |

## V. TEST RESULTS

We conducted a series of tests to evaluate the performance of this TIADC. Shown in Fig. 15 is the system under test. The sine wave signal from a RF signal generator Rohde & Schwarz SMA100A passes a coaxial Band Pass Filter (BPF) and is then fed to the TIADC.

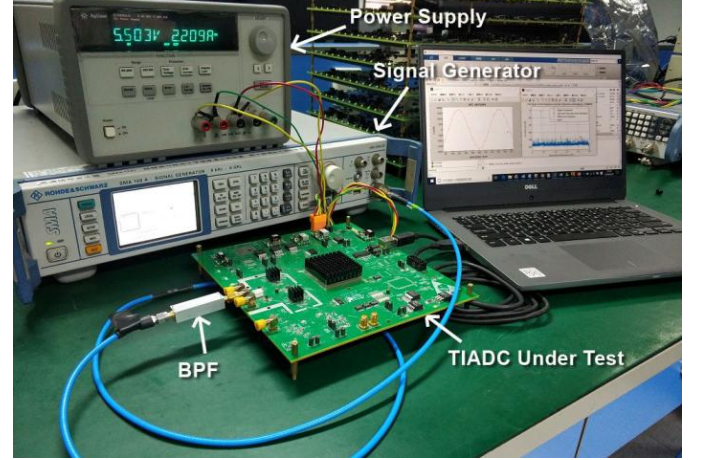

Fig. 15. System under test.

We performed dynamic analysis of the ADC system based on IEEE Std. 1241-2010 [43], and analyzed the data through the Fast Fourier Transform (FFT) and spectral averaging approach. Fig. 16 shows the frequency spectrum of the TIADC output with a 648 MHz sinusoid input signal, before and after correction. As we can observe, the distortion due to mismatch errors is suppressed by the correction algorithm.

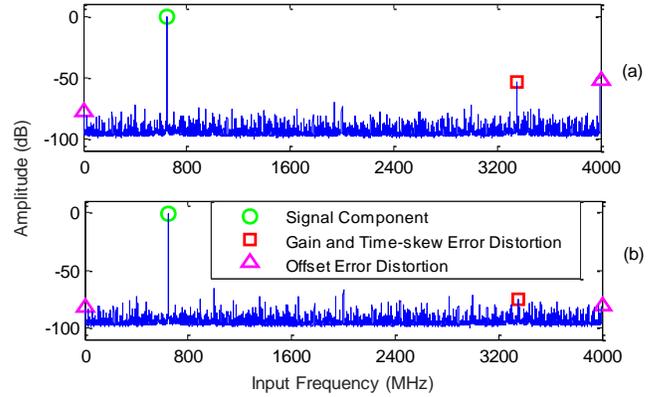

Fig. 16. Typical frequency spectrum. (a) before correction; (b) after correction.

We changed the input signal frequency, and tested the ENOB performance over a wide frequency range from 30 MHz to 1.6 GHz. In this frequency range, only one set of correction filters are employed, and their coefficients are kept constant.

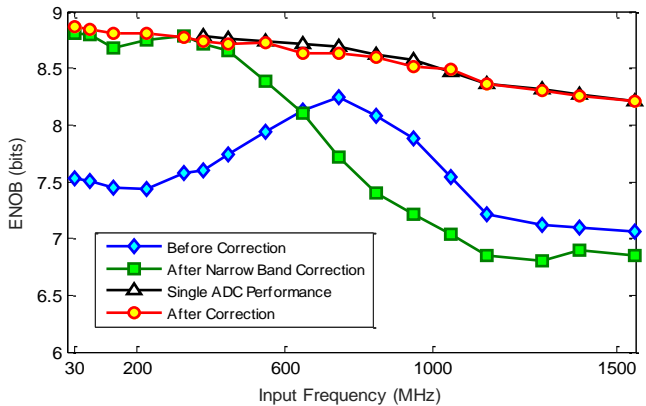

Fig. 17. ENOB performance test results.

As shown in Fig. 17, without correction the ENOB of the TIADC is deteriorated severely due to mismatch errors. With the wide band correction algorithm applied, the data can be corrected in real time. After correction, the ENOB is greatly enhanced to be better than 8.5 bits (<800 MHz) and 8 bits from 800 MHz to 1.6 GHz, almost the same as that of a single ADC chip according to its datasheet [29], which indicates that this correction algorithm performs well. Comparatively, we also plot the ENOB with the narrowband correction applied. As one can see, the narrowband correct only takes effect in a narrow frequency range, and the ENOB deteriorates fast with high input frequency, which corresponds just to the frequency range where the mismatch errors vary significantly.

To observe more directly the effect of the wide band correction method, we also conducted tests using the input signal combined from two sine wave signals with different frequencies. The test bench is shown in Fig. 18.

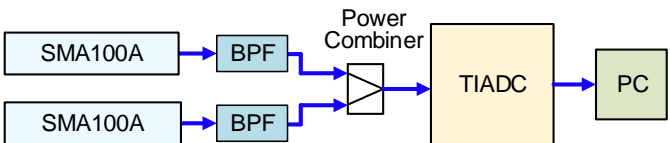

Fig. 18. Test bench using the input signal combined from two sine wave signals.

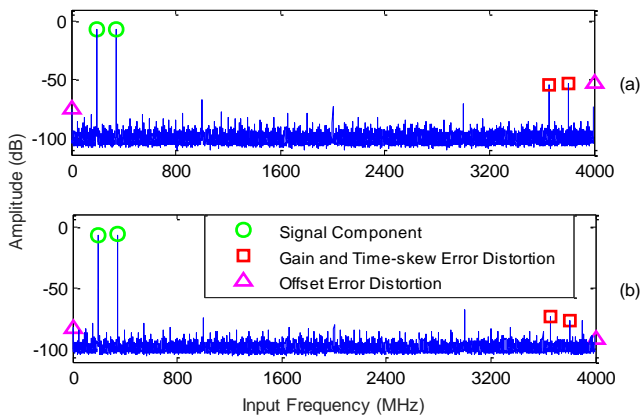

Fig. 19. Frequency spectrum with two input signal frequencies. (a) before correction; (b) after correction.

Fig. 19 shows the freqeuncy spetrum before and after correction, and it indicates that the interleaving spurious artifacts caused by gain and time skew errors is suprressed by more than 20 dB.

## VI. CONCLUSION

A 8-Gsps 12-bit TIADC is designed and tested. Based on numeric calculation of the FIR filter coefficient, a wide band correction algorithm is achieved and implemented as real-time processing logic in an FPGA. Test results indicate that this correction algorithm can effectively enhance the ENOB of the TIADC to be better than 8.5 bits from 30 MHz to 800 MHz and 8 bits from 800 MHz to 1.6 GHz, which is almost the same with that of the ADC chip.